\documentclass{iau} 
\usepackage{graphicx}
\pdfoutput=1 
\pubyear{2023}
\volume{373}
\jname{Resolving the Rise and Fall of Star Formation in Galaxies}
\editors{Kim W.-T. \& Wong T., eds.}

\title[JD 11.~~Star formation history for IC 10] 
{Star formation history for the starburst dwarf galaxy in the Local Group, IC 10}

\author[Mahtab Gholami et al.]   
{Mahtab Gholami$^1$, Atefeh Javadi$^1$, Jacco Th. van Loon$^2$, Habib Khosroshahi$^1$, \and Elham Saremi$^{1,3,4}$}

\affiliation{$^1$School of Astronomy, Institute for Research in Fundamental Sciences (IPM), Tehran, 19568-36613, Iran \\ email: {\tt mhtab.gholami@gmail.com} \\[\affilskip]
	$^2$Lennard-Jones Laboratories, Keele University, ST5 5BG, UK \\
	$^3$Instituto de Astrof{\`i}sica de Canarias, C/ V{\`i}a L{\`a}ctea s/n, 38205 La Laguna, Tenerife, Spain \\
	$^4$Departamento de Astrof{\`i}sica, Universidad de La Laguna, 38205 La Laguna, Tenerife, Spain}

\begin{document}
\maketitle
\begin{abstract}
IC 10 as a starburst dwarf galaxy in the Local Group (LG) has a large population of newly formed stars that are massive and intrinsically very bright in comparison with other LG galaxies. Using the Isaac Newton Telescope (INT) with the Wide Field Camera (WFC) in the i-band and V-band, we performed an optical monitoring survey to identify the most evolved asymptotic giant branch stars (AGBs) and red supergiant stars (RSGs) in this star-forming galaxy, which can be used to determine the star formation history (SFH). The E(B - V) as an effective factor for obtaining the precise magnitude of stars is measured for each star using a 2D dust map (SFD98) to obtain a total extinction for each star in both the i-band and V-band. We obtained the photometric catalog for 53579 stars within the area of 0.07 deg$^{2}$ (13.5 kpc$^{2}$), of which 762 stars are classified as variable candidates after removing the foreground stars and saturated ones from our catalog. To reconstruct the SFH for IC 10, we first identified 424 long-period variable (LPV) candidates within the area of two half-light radii (2r$_{h}$) from the center of the galaxy. We estimated the recent star formation rate (SFR) at $\sim$ 0.32 M$_{\odot}$ yr$^{-1}$ for a constant metallicity Z = 0.0008, showing the galaxy is currently undergoing high levels of star formation. Also, a total stellar mass of 0.44 $\times$ 10$^{8}$ M$_{\odot}$ is obtained within 2r$_{h}$ for that metallicity.
\end{abstract}

\firstsection 
\section{Introduction}
A dwarf irregular galaxy, IC 10 (UGC 192) is one of the most interesting systems and is the nearest starburst dwarf located within the LG (e.g., \cite[Gerbrandt et al. 2015]{Gerbrandt15}). The existence of young, intermediate, and old-age stellar populations has been studied in IC 10 (e.g., \cite[Massey \& Armandroff 1995]{Massey95}, \cite[Sakai et al. 1999]{Sakai99}, \cite[Borissova et al. 2000]{Borissova00}) in which the highest density of Wolf-Rayet stars indicates the recent burst of star formation (e.g., \cite[Cosens et al. 2022]{Cosens22}). On the other hand, the light of the galaxy is affected by gas-dust clouds of the Milky Way (MW) galaxy because IC 10 is located in the Galactic plane (l = 118$^{\circ}$.9, b = -3$^{\circ}$.3). Therefore, studying this galaxy is restricted due to the strong extinction of MW (\cite[Richer et al. 2001]{Richer01}). 
The greatest readily available tracers of stellar populations are evolved stars with high luminosity, such as AGBs and RSGs (\cite[Javadi et al. 2011b]{Javadi11b}). For the AGBs and RSGs which are at the final stage of their evolution, their luminosity would be directly proportional to their birth mass, and they are exceptional objects for reconstructing the SFH for a galaxy and tracing stellar populations from as recently formed as 10 Myr to as ancient as 10 Gyr (\cite[Javadi et al. 2011b]{Javadi11b}).
In this paper, we conducted an optical survey of the majority of dwarf galaxies in the LG between June 2015 and October 2017 with the 2.5-m INT/WFC over nine epochs (\cite[Saremi et al. 2020]{Saremi20}). We identified the LPV candidates of IC 10, then, based on their luminosity distribution, we determined the SFH using the method that we successfully applied previously to a number of LG galaxies in our survey (e.g., \cite[Javadi et al. 2017]{Javadi17}, \cite[Hashemi et al. 2018]{Hashemi18}, \cite[Saremi et al. 2021]{Saremi21}, \cite[Navabi et al. 2021]{Navabi21}, \cite[Parto et al. 2023]{Parto23}, \cite[Abdollahi et al. 2023]{Abdollahi23}); Here, we describe our method for reconstructing the SFH of IC 10 using our identified LPV candidates.
\firstsection
\section{Identification of LPV candidates}
A photometry catalog was obtained for 53579 stars in the entire area of CCD4 (0.07 deg$^{2}$ ($13.5$ kpc$^{2}$)) using the {\sc DAOPHOT}/{\sc ALLSTAR} software package (\cite[Stetson 1987]{Stetson87}). We estimated amount of foreground contamination for the galaxy using the stars in common between our catalog and the Gaia Early Data Release 3 catalog (Gaia EDR3) (\cite[Lindegren et al. 2021]{Lindegren21}). Then, we determined the foregrounds based on the parallax and proper motion measurements of stars. We estimated the percentage of contamination of our catalog by foreground stars $\sim$ 7\,\%. Finally, we removed them from our catalog for further analysis. As a result, 762 stars are identified as variable stars. Among them, 424 stars are identified as LPV candidates due to large amplitude variability, mostly more than 0.20 mag within the area of 2r$_{h}$ centered on the galaxy. Moreover, we applied the extinction correction from SFD98 dust map (\cite[Schlegel, Finkbeiner \& Davis 1998]{Schlegel98}) in both i-band (i$_{0}$) and V-band (V$_{0}$) magnitudes for all stars in our catalog.
\firstsection
\section{Star formation history for IC 10}
\begin{figure}[h!]
\begin{center}
 \includegraphics[width=3.4in]{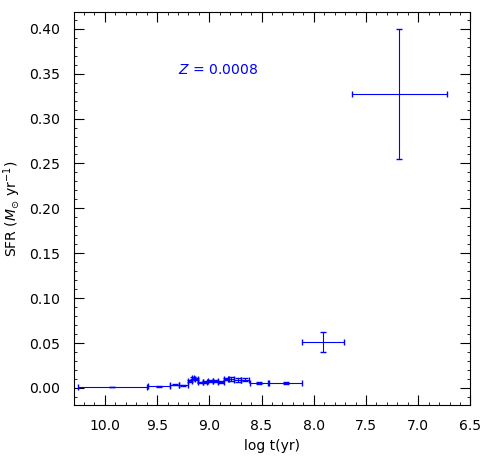}
\caption{SFH of IC 10 for constant metallicity of Z = 0.0008. The horizontal and vertical error bars represent the age range and the statistical errors, respectively in each bin derived for LPV candidates within 2r$_{h}$ area.}
 \label{fig:sfr8}
\end{center} 
\end{figure}

We derived the SFH of IC 10 following the same method applied to other LG dwarf galaxies described in previous papers (e.g., \cite[Saremi et al. 2021]{Saremi21}, \cite[Navabi et al. 2021]{Navabi21}). The SFH has been estimated by calculation of the SFR, $\xi$, in M$_{\odot}$yr$^{-1}$ based on the LPV candidates. The value of SFR, $\xi$, shows the amount of gas mass that was converted into numerous stars per year. The number of newborn stars can be determined from their total mass, which is measured by using an initial mass function (IMF) (\cite[Kroupa 2001]{Kroupa01}) in which the minimum and maximum ranges of stellar mass have been determined as 0.02 M$_{\odot}$ and 200 M$_{\odot}$, respectively. The $\xi(t)$ in each age bin (dt), is obtained using the following equation in which the dn$^{'}$ is the number of LPV candidates in each age bin.
\begin{equation}\label{eq:eq1}
\xi(t) = \frac{dn^{'}(t)}{\delta t} \frac{\int_{min}^{max} f_{IMF}(m)m\,dm}{\int_{m(t)}^{m(t+dt)}f_{IMF}(m)\,dm}
\end{equation}
In the equation \ref{eq:eq1}, the range of the stellar mass for LPV candidates is defined between m(t) and m(t+dt) while the $\delta$t is the time duration of pulsation of those stars. The f$_{IMF}$ is defined using f$_{IMF}$ = Am$^{-\alpha}$ where A and $\alpha$ are recognized as the constant of normalization and the mass range dependence, respectively (\cite[Kroupa 2001]{Kroupa01}). Fig.\,\ref{fig:sfr8} shows the SFH for a constant metallicity of Z = 0.0008 over the time period of 5 Myr to 13.8 Gyr ago. The horizontal and vertical error bars indicate the age range and statistical errors in each bin, respectively. The statistical errors for each age bin are also calculated assuming Poisson statistics. As can be seen in the Fig.\,\ref{fig:sfr8}, the figure for SFH shows a peaking around 10 Myr ago (log\,t $\sim$ 7) at 0.32 $\pm$ 0.07 M$_{\odot}$ yr$^{-1}$ which indicates high star-forming activity in IC 10 in the recent years. Also, there is another major peak coming up at 0.05 $\pm$ 0.07 M$_{\odot}$ yr$^{-1}$ at around 80 Myr ago. It explains the high star-forming activity in that bin followed by a significant increase in recent years. Also, the total stellar mass within 2r$_{h}$ of the galaxy is obtained (0.44 $\pm$ 0.08) $\times$ 10$^{8}$ M$_{\odot}$.
\firstsection 
\section{Conclusion}
We have conducted an optical monitoring survey of the majority of dwarf galaxies in the LG with WFC at INT. Based on our obtained 424 LPV candidates within 2r$_{h}$, we explain our method to reconstruct the SFH of the starburst dwarf, IC 10. For each age bin, we calculated the SFR with a constant metallicity Z = 0.0008; the SFH is prominent at recent times at 0.32 $\pm$ 0.07 M$_{\odot}$ yr$^{-1}$ around 10 Myr ago. It shows the current intense star formation activity of IC 10.
\def\apj{{ApJ}}    
\def\nat{{Nature}}    
\def\jgr{{JGR}}    
\def\apjl{{ApJ Letters}}    
\def\aap{{A\&A}}   
\def\mnras{{MNRAS}}
\def\aj{{AJ}}
\let\mnrasl=\mnras
\firstsection

\end{document}